\documentclass{JHEP3}
\usepackage[dvips]{graphicx}
\usepackage{amsmath,amssymb}

\newcommand{\be}{\begin{equation}}
\newcommand{\ee}{\end{equation}}
\newcommand{\bea}{\begin{eqnarray}}
\newcommand{\eea}{\end{eqnarray}}

\newcommand{\mat}{\left ( \begin{array}{cc}}
\newcommand{\emat}{\end{array} \right )}

\newcommand{\bfe}{{\mathbf e}}

\newcommand{\bfm}{{\mathbf m}}
\newcommand{\bfn}{{\mathbf n}}

\newcommand{\bphi}{{\overline{\phi}}}

\newcommand{\cF}{{\cal F}}
\newcommand{\cI}{{\cal I}}
\newcommand{\cL}{{\cal L}}
\newcommand{\cN}{{\cal N}}

\newcommand{\vchi}{{\vec{\chi}}}
\newcommand{\vH}{{\vec{H}}}
\newcommand{\vPhi}{{\vec{\Phi}}}
\newcommand{\vPsi}{{\vec{\Psi}}}

\newcommand{\Z}{\mathbb{Z}}

\newcommand{\nn}{\nonumber}

\newcommand{\Tr}{{\rm Tr}\,}

\newcommand{\e}{\epsilon}

\title{Relations among Supersymmetric Lattice Gauge Theories 
via Orbifolding
}
\author{Poul H. Damgaard and So Matsuura\\ The Niels Bohr Institute, 
The Niels Bohr International Academy,
Blegdamsvej 17, DK-2100 Copenhagen, Denmark}
\abstract{We show how to derive Catterall's supersymmetric lattice
gauge theories directly from 
the general principle of orbifolding 
followed by a variant of the usual deconstruction. 
These theories are forced to be complexified due to a clash
between charge assignments under U(1)-symmetries and lattice 
assignments in terms of scalar, vector and tensor components for
the fermions. 
Other prescriptions for how to discretize the theory 
follow automatically by orbifolding and deconstruction. 
We find that Catterall's complexified
model for the two-dimensional $\cN=(2,2)$ theory 
has two independent
preserved supersymmetries.
We comment on consistent truncations to lattice theories
without this complexification and with the correct continuum limit.  
The construction of lattice theories this way is general, 
and can be used to derive new supersymmetric lattice theories
through the orbifolding procedure. 
As an example, we apply the prescription to topologically twisted 
four-dimensional $\cN=2$ supersymmetric Yang-Mills theory. 
We show that a consistent truncation is closely related
to the lattice formulation previously given by Sugino. 
}
\begin{document}

\section{Introduction}

Recently, there has been a rapid series of developments in the lattice
construction of supersymmetric gauge theories 
%Kaplan
\cite{Kaplan:2002wv}--\nocite{Cohen:2003xe}%
\nocite{Cohen:2003qw}\nocite{Kaplan:2005ta}\nocite{Damgaard:2007be}%
%Catterall
\nocite{Catterall:2003wd}\nocite{Catterall:2004np}\nocite{Catterall:2005fd}%
%Sugino
\nocite{Sugino:2003yb}\nocite{Sugino:2004qd}%
\nocite{Sugino:2004uv}\nocite{Sugino:2006uf}%
%Kawamoto
\nocite{Suzuki:2007jt}\nocite{D'Adda:2004jb}\cite{D'Adda:2005zk}.%
\footnote{
Possible difficulties with these formulations are discussed in 
\cite{Giedt:2003ve}\nocite{Bruckmann:2006ub}--\cite{Bruckmann:2006kb}.
}
These lattice formulations share one common feature; 
there is at least one preserved nilpotent scalar supercharge $Q$, 
which is a part of the 
original supersymmetry generators of the continuum theories. 
Roughly speaking, 
which supercharge is chosen to be preserved on the lattice
determines the lattice formulation. 
This way of defining supersymmetric lattice gauge theories lies
very close to the way topological field theories are defined in the
continuum based on BRST symmetry $Q$. As is well-known,
such topological field theories in the continuum can alternatively be 
viewed as twistings of ordinary field theories with space-time 
supersymmetry. It is therefore very natural to define lattice gauge
theories with some remnant(s) of supersymmetry by means of the
same procedure. After an ``untwisting'' on the lattice one can
define physical observables which hopefully will not suffer
from the usual fine tuning problems of other approaches to lattice
supersymmetry.

In refs.~%
\cite{Kaplan:2002wv}--\nocite{Cohen:2003xe}%
\nocite{Cohen:2003qw}\cite{Kaplan:2005ta},  %\cite{Damgaard:2007be}, 
a systematic way to generate lattice structure from 
a matrix theory (the ``mother theory'') has been presented.
Here the preserved supercharge is one component of the original
supersymmetry generators in general.% 
\footnote{For a discussion of the
relations between these lattice theories and topological field theories, 
see ref.~\cite{Unsal:2006qp}.}
In this formulation, the space-time lattice itself is generated by 
orbifolding followed by deconstruction \cite{Arkani-Hamed:2001ca}, 
and the dimensionality is
determined by the number of the maximal global $U(1)$ symmetries of the
mother theory. 
Therefore, possible lattice theories generated from a given mother
theory are restricted. 
A classification of 
orbifolded theories with up to eight supercharges has recently been
given in \cite{Damgaard:2007be}. 

Among 
alternative lattice formulations of supersymmetric gauge theories are 
those due to 
Catterall \cite{Catterall:2003wd}--%
\nocite{Catterall:2004np}\cite{Catterall:2005fd} 
and Sugino 
\cite{Sugino:2003yb}--\nocite{Sugino:2004qd}%
\nocite{Sugino:2004uv}\cite{Sugino:2006uf}, 
both of which preserve the BRST charge of a topologically twisted
supersymmetric gauge theory \cite{Witten:1988ze}. 
The idea of 
both of these formulations is to write down lattice actions that
are $Q$-exact at fixed lattice spacing. 
Although they thus seem to be close to each other in spirit, 
they appear very different in detail at first sight. 
One surprising feature of Catterall's formulation is that it seems
to require a complexification of fields in order to preserve both
gauge invariance and some remnant of supersymmetry on the lattice. 
After constructing the lattice action for the complexified fields, 
the path-integral has been restricted to the ``real line'' in actual
simulations. 
By this restriction, however, one breaks both gauge
symmetry and the remnant of supersymmetry.
Nevertheless, simulations done ``on the real line'' \cite{Catterall:2006jw} 
seem to indicate a surprisingly good approximation to the supersymmetry
one hopes to recover in the continuum. 
Sugino's formulation, on the other hand, does not need this
complexification. 
Yet, both are supposed 
to be discretizations of the corresponding topological field
theories in the continuum.  
For numerical simulations for Sugino's model, see \cite{Suzuki:2007jt}.

Very recently, in a very interesting paper \cite{Takimi:2007nn}, 
Takimi has shown that the theories of Sugino and the complexified 
theories of Catterall are indeed connected. 
More precisely, the degrees of freedom 
of Catterall's complexified lattice theory for two-dimensional 
$\cN=(2,2)$ supersymmetric gauge theory can be reduced in
a manner consistent with both gauge symmetry and supersymmetry. 
The resulting theory is, after some field redefinitions, 
very closely related to Sugino's lattice formulation.

In this paper, we wish to understand 
Catterall's theories from the orbifolding procedure. 
In ref.~\cite{Damgaard:2007be}, 
we derived what we believe is the complete classification of 
orbifolded theories with up to eight supercharges 
and none of these 
theories seemed to include those of Catterall. Is the orbifolding
technique not the most general way to generate such supersymmetric
lattice theories? Or was the classification incomplete?
As we shall show, the answer lies in the restrictions one 
imposes on oneself if one insists on a particular assignment of
fields on the lattice. In particular, the crucial part is the
way one insists on identifying fields transforming irreducibly
under Lorentz transformations. If one {\em beforehand} insists on 
scalars, vectors and tensors in the continuum being represented
by site variables, links and corner variables, respectively, then
one may run into clashes with the orbifolding technique. This is
because the assignment of $U(1)$-charges (some of which are
subsets of Lorentz symmetries) is in a one-to-one
correspondence with the generation of the lattice itself. In the
case of Catterall's prescription, these $U(1)$-charges do not
match those required for the lattice assignments that are
being insisted upon.
The apparently only way out is to {\em complexify}\footnote{For 
another way to make connection with the orbifolding procedure, 
see ref. \cite{Takimi:2007nn}.}. As we shall
show, this can be done so that it introduces
just the right amount of additional $U(1)$-symmetries. The price
one pays is that one is not considering the right theory anymore,
but a complexified one.

Having understood that this is the way to generate the
complexified supersymmetric theories according to Catterall's 
prescription, it
is now a simple matter to generalize this to many other theories.
In particular, there is apparently no deeper need to tie oneself
up to theories that admit a complete description in terms
of Dirac-K\"{a}hler fields. If one allows oneself to complexify,
many other theories are possible. We shall illustrate this by
showing how to generate a complexified version of ${\cN}=2$
supersymmetric lattice gauge theory in four dimensions by a
combination of complexification and orbifolding. As for
Catterall's examples, going to the real line breaks both gauge 
symmetry and the last remnant of lattice supersymmetry\footnote{But 
if the numerical experience of ref.~\cite{Catterall:2006jw} 
holds here too,
this may be a quite good approximation to such
supersymmetric lattice gauge theories.}.
Instead, we demonstrate that we can truncate to fewer degrees of
freedom while preserving both gauge symmetry
and supersymmetry, just as was done in ref. \cite{Takimi:2007nn}. 
The obtained theory is again essentially, up to a few
additional terms, equal to Sugino's formulation of 
four-dimensional $\cN=2$ supersymmetric Yang-Mills theory
\cite{Sugino:2003yb}. 

Our paper is organized as follows. In section 2 we show how to derive
Catterall's complexified (2,2)-model by combining orbifolding and 
complexification. 
Surprisingly, we find that this theory actually is
invariant under two different scalar supercharges $Q_{\pm}$, not
just one as previously believed. The two charges $Q_+$ and
$Q_-$ can be viewed as BRST and anti-BRST charges, respectively,
and the action is exact in both of them. 
We discuss the problems that arise if one tries to project the resulting
complexified 
theory onto the real line: loss of lattice supersymmetry in both
the action and the measure (and the combination of the two).
In section 3 we comment on the recent observation by Takimi
\cite{Takimi:2007nn} 
of a consistent truncation of Catterall's complexified
model that turns out to be closely related to Sugino's
\cite{Sugino:2004qd}.
Because of the existence of two independently conserved 
supersymmetry charges, we can consider the same type of truncation
based on the other supersymmetry charge. As it turns out, it
yields the same action, up to trivial changes of
conventions. In section 4 we discuss possible generalizations of 
Catterall's 
complexified models that can be constructed by orbifolding. 
This includes many supersymmetric theories that could not be
derived by orbifolding in the conventional way, including
${\cal N}=2$ supersymmetric Yang-Mills theory in four dimensions.
The challenge is then to find either consistent truncations,
or truncations that, although they may break all 
supersymmetries, may still yield supersymmetric field theories
in the continuum without the need of fine tuning. 
We show that we obtain a theory very closely related to 
Sugino's formulation of four-dimensional
$\cN=2$ supersymmetric Yang-Mills theory \cite{Sugino:2003yb} by 
one particular truncation, followed by field redefinitions.
In section 5 we present our conclusions.

\section{Catterall's Construction from Orbifolding}

In this section we show how to obtain Catterall's complexified lattice 
gauge theories by the orbifolding procedure of 
refs.~\cite{Kaplan:2002wv}--\nocite{Cohen:2003xe}%
\nocite{Cohen:2003qw}\nocite{Kaplan:2005ta}\cite{Damgaard:2007be}.
In particular, we show that the discretization prescription given 
in \cite{Catterall:2004np} can be clearly understood by this procedure.
To be definite, we concentrate on the lattice theory for two-dimensional 
$\cN = (2,2)$ supersymmetry in the continuum limit. 
As part of our derivation, we will
also show that there is an additional, hidden, (anti-)BRST-like symmetry in 
Catterall's model. 

\subsection{Derivation of Catterall's action by the orbifolding procedure}
As usual with orbifolding technique, we begin with a ``mother theory'', 
here a matrix model obtained by dimensional reduction of
$\cN=1$ supersymmetric Yang-Mills theory in four-dimensional 
Euclidean space-time,
\be
S = \frac{1}{g^2}{\rm Tr}\left(-\frac{1}{4}[v_{\alpha},v_{\beta}]^2 
+ \frac{i}{2}
\bar{\Psi}\Gamma_\alpha[v_\alpha,\Psi]\right), 
\qquad (\alpha,\beta=0,\cdots,3)
\label{Smother0}
\ee
where $\Gamma_\alpha$ are $SO(4)$ Dirac matrices, 
$v_\alpha$ are $kN^2\times kN^2$ hermitian matrices, 
$\Psi$ is a four-component fermion and $\bar\Psi\equiv \Psi^T C$ with 
the charge conjugation matrix $C$ satisfying 
$C^{-1}\Gamma_{\alpha}C = -\Gamma_{\alpha}^T$.
Following \cite{Cohen:2003xe}, 
we choose the notation of the $\gamma$-matrices and the
charge conjugation matrix as 
\be
\Gamma_{\alpha} = \mat 0  & \sigma_{\alpha} \\
     \bar{\sigma}_{\alpha} &  0 \emat, \qquad 
C = \mat i\tau_2 & 0 \\
                                0 & -i\tau_2 \emat,
\label{Q=4 mother}
\ee
with $\sigma_{\alpha} = ({\mathbf 1}, -i\tau_i)$ and 
$\bar{\sigma}_{\alpha} = ({\mathbf 1},i\tau_i)$ where $\tau_i$
($i=1,2,3$) are Pauli matrices.
Our purpose in this section is to obtain a lattice regularization of 
topologically twisted two-dimensional $\cN=(2,2)$ supersymmetric gauge
theory. To this end, we rearrange the fields so that the symmetry of the
two-dimensional theory becomes manifest:
\begin{align}
 v_0 &\equiv A_1, \quad v_3 \equiv -A_2, \quad 
 v_1+iv_2 \equiv i\phi, \quad 
 v_1-iv_2 \equiv -i\bphi, \nn \\
 \Psi^{(1)} &\equiv \left(\begin{matrix}
-i\chi_{12}-\frac{1}{2}\eta \\
\psi_1-i\psi_2 \end{matrix}\right), \quad 
 \Psi^{(2)} \equiv \left(\begin{matrix}
-i\chi_{12}+\frac{1}{2}\eta \\
\psi_1+i\psi_2
\end{matrix}\right), 
\end{align}
where we have set $\Psi^T\equiv\left(\Psi^{(1)T},\Psi^{(2)T}\right)$.
Then the action (\ref{Q=4 mother}) can be rewritten as 
\begin{align}
 S 
 %&= \frac{1}{g^2}\Tr\Bigl\{
% -\frac{1}{4}[A_\mu,A_\nu]^2 
% -\frac{1}{2} [A_\mu,\phi][A_\mu,\bphi] 
% -\frac{1}{8}[\phi,\bphi]^2  
% +\eta [A_\mu,\psi_\mu]\nn \\
% &\hspace{2cm}
% -\chi_{\mu\nu}\left([A_\mu,\psi_\nu]-[A_\nu,\psi_\mu]\right)
% +\frac{1}{4}\eta[\phi,\eta] 
% -\psi_\mu \left[\bphi,\psi_\mu\right]
% +\frac{1}{2}\chi_{\mu\nu}\left[\phi,\chi_{\mu\nu}\right]
%\Bigr\} \nn \\
&=\frac{1}{g^2}\Tr\Bigl\{
-B_{\mu\nu}^2 +iB_{\mu\nu}[A_\mu,A_\nu] 
 -\frac{1}{2} [A_\mu,\phi][A_\mu,\bphi] 
 +\frac{1}{8}[\phi,\bphi]^2  
 -i\eta [A_\mu,\psi_\mu]\nn \\
 &\hspace{2cm}
 -i\chi_{\mu\nu}\left([A_\mu,\psi_\nu]-[A_\nu,\psi_\mu]\right)
 -\frac{i}{4}\eta[\phi,\eta] 
 +i\psi_\mu \left[\bphi,\psi_\mu\right]
 -\frac{i}{2}\chi_{\mu\nu}\left[\phi,\chi_{\mu\nu}\right]
\Bigr\}
\label{twisted N=(2,2)}
\end{align} 
where $\chi_{12}=-\chi_{21}$ 
and we have introduced an auxiliary field $B_{\mu\nu}=-B_{\nu\mu}$. 
As discussed in \cite{Catterall:2004np}, we should regard $\phi$ and 
$\bphi$ as independent hermitian matrices rather than complex
conjugate. 
In the expression (\ref{twisted N=(2,2)}), a scalar supersymmetry 
(equivalently, a BRST symmetry) is manifest, 
and we can rewrite the action in a $Q$-exact form as 
\begin{align}
 S = \frac{1}{g^2} \Tr Q \Bigl\{
 -\chi_{\mu\nu}\left(B_{\mu\nu}-i[A_\mu,A_\nu]\right) 
 +i \psi_\mu[A_\mu,\bphi]
 +\frac{i}{4}\eta[\phi,\bphi]
\Bigr\},  
\label{Q-exact mother}
\end{align}
where $B_{\mu\nu}$ is a auxiliary field and $Q$ is the BRST charge 
which acts on the fields as 
\begin{alignat}{2}
 &Q A_\mu = \psi_\mu, &\quad &Q \psi_\mu = \frac{i}{2}[A_\mu,\phi], \nn
 \\
\label{original BRST charge}  
 &Q \bphi = \eta, &\qquad &Q\eta = -\frac{i}{2}[\phi,\bphi], \\ 
 &Q \chi_{\mu\nu} = B_{\mu\nu}, &\quad 
 &Q B_{\mu\nu} = -\frac{i}{2}[\phi,\chi_{\mu\nu}], \qquad 
 Q\phi = 0. \nn
\end{alignat}
One can easily show that $Q^2=\delta_{-\phi/2}$, where $\delta_\theta$ is
the gauge transformation with a parameter $\theta$. 
Thus, $Q$ is nilpotent up to gauge transformations.

Next, we would like to derive a lattice theory from the mother theory
(\ref{Q-exact mother}) using orbifolding and deconstruction while
preserving the BRST charge $Q$. 
To do so, we must first specify two $U(1)$ symmetries to create a
two-dimensional lattice. (For details, 
see \cite{Kaplan:2002wv}--\nocite{Cohen:2003xe}%
\nocite{Cohen:2003qw}\nocite{Kaplan:2005ta}\cite{Damgaard:2007be}.)
In our case, we must demand of these $U(1)$ symmetries that 
the BRST operator $Q$ has zero charges and all fields have definite
charges so that the action (\ref{Q-exact mother}) has 
zero charge.  
However, we immediately see that it is impossible. 
In fact, since the gauge fields $A_\mu$ should become link variables, 
they must have non-zero charges. 
Then, from the BRST transformation (\ref{original BRST charge}), 
$\psi_\mu$ must have the same $U(1)$ charges as
$A_\mu$, while $\{\phi,\bphi,\eta\}$ should have zero charges. 
Under this condition, the $U(1)$ charges of the second term 
of (\ref{complex mother}) cannot be zero; 
it is impossible to assign non-vanishing definite $U(1)$ charges to
the fields. 
%We cannot define non-trivial $U(1)$ charges for $A_\mu$ since
%they are hermitian matrices by definition. 
%Moreover, by insisting on
%the assignments of charges to the different fields (scalar variables like
%$\eta$ should have vanishing charges, vector variables $\psi_{\mu}$
%should have charges corresponding to the two directions generated
%by the orbifolded lattice, and so on), one immediately sees that no
%valid charge combination exists. 
This is consistent 
with our earlier result \cite{Damgaard:2007be} 
that the two-dimensional lattice
theory constructed by orbifolding from the mother theory (\ref{Smother0}) is 
unique, and coincides with the one given in \cite{Cohen:2003xe}. 

In order to avoid this problem, we extend, as is done in 
ref.~\cite{Catterall:2004np}, all fields except 
$\phi$ and $\bphi$ to complex matrices, and we change simultaneously the action 
(\ref{Q-exact mother}) as follows:
\begin{align}
 S = \frac{1}{2g^2} \Tr Q_+ \Bigl\{
 &\chi_{\mu\nu}^\dagger\Bigl(-B_{\mu\nu}+i[A_\mu,A_\nu]\Bigr) 
 +\chi_{\mu\nu}\left(-B_{\mu\nu}^\dagger+i[A_\mu^\dagger,A_\nu^\dagger]\right)
 \nn \\
 &+i \psi_\mu^\dagger [A_\mu,\bphi] +i \psi_\mu [A_\mu^\dagger,\bphi]
 %- \frac{1}{4}\bareta[\phi,\bphi] 
 +\frac{i}{4}\eta_+[\phi,\bphi] +\frac{1}{2}\eta_- d
 \Bigr\}, 
 \label{complex mother}
 \end{align}
where $A_\mu^\dagger$, $B_{\mu\nu}^\dagger$ and $\psi_\mu^\dagger$ are 
hermitian conjugate of $A_\mu$, $B_{\mu\nu}$ and $\psi_\mu$,
respectively, $\eta_+$ and $\eta_-$ are independent hermitian matrices 
and $d$ is a hermitian auxiliary field. 
The BRST charge $Q_+$ is a natural extension of $Q$ 
in (\ref{original BRST charge}) which act to the fields as 
\begin{alignat}{2}
 &Q_+ A_\mu = \psi_\mu, &\qquad &Q_+ \psi_\mu =  \frac{i}{2}[A_\mu,\phi], \nn \\
 &Q_+ A_\mu^\dagger = \psi_\mu^\dagger, &\quad 
  &Q_+ \psi_\mu^\dagger =  \frac{i}{2}[A_\mu^\dagger,\phi], \nn \\
 &Q_+ \bphi = \eta_+, &\quad &Q_+ \eta_+ = -\frac{i}{2}[\phi,\bphi], \nn \\
\label{BRST charge pre 1}  
 &Q_+ d = -\frac{i}{2}[\phi,\eta_-], &\quad &Q_+ \eta_- = d, \\
 &Q_+ \chi_{\mu\nu} = B_{\mu\nu}, &\quad 
 &Q_+ B_{\mu\nu} = -\frac{i}{2}[\phi,\chi_{\mu\nu}], \nn \\
 &Q_+ \chi_{\mu\nu}^\dagger = B_{\mu\nu}^\dagger, &\quad 
 &Q_+ B_{\mu\nu}^\dagger = -\frac{i}{2}[\phi,\chi_{\mu\nu}^\dagger],
 \qquad Q_+ \phi = 0. \nn 
\end{alignat}
The charge $Q_+$ is nilpotent up to gauge
transformations, just as was the original $Q$. 
It is easy to see that (\ref{complex mother}) returns to the original
form (\ref{twisted N=(2,2)}) if we take $A_\mu^\dagger=A_\mu$, 
$B_{\mu\nu}^\dagger=B_{\mu\nu}$, $\psi_\mu^\dagger=\psi_\mu$, 
$d=\eta_-=0$ and $\eta_+=\eta$.

By the above extension, the action acquires extra $U(1)$ symmetries 
and the action is invariant under the transformation, 
\begin{equation}
 \Phi \to e^{iq_1\theta_1+iq_2\theta_2} \Phi,
 \qquad \left(\theta_1,\theta_2 \in [0,2\pi)\right)
\end{equation}
where $\Phi$ is a collective field content in the action 
(\ref{complex mother}),  
and the $U(1)$ charges $q_1$ and $q_2$ are given 
in Table \ref{charge table}. 
For the purpose of the future discussion, we introduce two vectors, 
\begin{equation}
 \bfe_1 \equiv\left(\begin{matrix}1\\0\end{matrix}\right), \quad 
 \bfe_2 \equiv\left(\begin{matrix}0\\1\end{matrix}\right). 
\end{equation}
\begin{table}[h]
\caption{The charge assignment for the complexified fields} 
\begin{center}
\begin{tabular}{c|cccccccccc}
 & $A_1$ & $A_2$ & $\phi$ & $\bphi$ & $B_{12}$ 
 & $\eta_+$ & $\eta_-$ & $\psi_1$ & $\psi_2$ & $\chi_{12}$ \\
\hline
$q_1$ & 1 & 0 & 0 & 0 & 1 & 0 & 0 & 1 & 0 & 1 \\
$q_2$ & 0 & 1 & 0 & 0 & 1 & 0 & 0 & 0 & 1 & 1
\end{tabular}
\end{center}
\label{charge table}
\end{table}
As discussed in \cite{Damgaard:2007be}, 
the orbifolded action is obtained by substituting the following 
expansion of the fields in (\ref{complex mother}):
\begin{alignat}{2}
 A_\mu &= \sum_{\bfn\in\Z_N^2} A_\mu(\bfn)\otimes E_{\bfn,\bfn+\bfe_\mu}, 
 &\quad 
 A_\mu^\dagger &= \sum_{\bfn\in\Z_N^2} A_\mu^\dagger(\bfn)\otimes 
 E_{\bfn+\bfe_\mu,\bfn} \nn \\
 \phi &= \sum_{\bfn\in\Z_N^2} \phi(\bfn)\otimes E_{\bfn,\bfn}, 
 &\quad
 \bphi &= \sum_{\bfn\in\Z_N^2} \bphi(\bfn)\otimes E_{\bfn,\bfn}, \nn \\
\label{orbifolding} 
B_{12} &= \sum_{\bfn\in\Z_N^2} B_{12}(\bfn)\otimes E_{\bfn+\bfe_1+\bfe_2,\bfn},
 &\quad 
 B_{12}^\dagger &= \sum_{\bfn\in\Z_N^2} B_{12}^\dagger(\bfn)\otimes 
 E_{\bfn,\bfn+\bfe_1+\bfe_2}, \\
 \eta_+ &= \sum_{\bfn\in\Z_N^2} \eta_+(\bfn)\otimes E_{\bfn,\bfn}, 
 &\quad
 \eta_- &= \sum_{\bfn\in\Z_N^2} \eta_-(\bfn)\otimes E_{\bfn,\bfn}, \nn \\
 \psi_\mu &= \sum_{\bfn\in\Z_N^2} \psi_\mu(\bfn)\otimes E_{\bfn,\bfn+\bfe_\mu}, 
 &\quad 
 \psi_\mu^\dagger &= \sum_{\bfn\in\Z_N^2} \psi_\mu^\dagger(\bfn)\otimes 
 E_{\bfn+\bfe_\mu,\bfn} \nn \\
 \chi_{12} &= \sum_{\bfn\in\Z_N^2} \chi_{12}(\bfn)\otimes E_{\bfn+\bfe_1+\bfe_2,\bfn},
 &\quad 
 \chi_{12}^\dagger &= \sum_{\bfn\in\Z_N^2} \chi_{12}^\dagger(\bfn)\otimes 
 E_{\bfn,\bfn+\bfe_1+\bfe_2}, \nn \\
 d &= \sum_{\bfn\in\Z_N^2} d(\bfn)\otimes E_{\bfn,\bfn}, \nn
\end{alignat}
where $E_{\bfm,\bfn}$ $(\bfm=(m_1,m_2),\ \bfn=(n_1,n_2))$ is an
$N^2\times N^2$ matrix defined by 
\begin{equation}
 E_{\bfm,\bfn} \equiv E_{m_1,n_1}\otimes E_{m_2,n_2}. \quad 
\Bigl( \left(E_{i,j}\right)_{kl}=\delta_{ik}\delta_{jl}, \quad
i,j,k,l=1,\cdots,N \Bigr)
\end{equation}

%According to the procedure to create a lattice theory, 
Furthermore, in the standard method of deconstruction, we search for flat directions,
and use these to shift appropriate combinations of fields in order to generate
kinetic terms. Here we wish to shift the fields $A_\mu$ and $A_\mu^\dagger$ 
with the amount of $1/a$ in order to introduce such kinetic terms
for the gauge potentials, and by gauge symmetry, all other fields with
non-trivial couplings to these gauge potentials.
Instead of this shift operation, however, we could replace 
$A_\mu(\bfn)$ and $A_\mu^\dagger(\bfn)$ as \cite{Unsal:2006qp} 
\begin{align}
 A_\mu(\bfn)&\to \frac{1}{ia} e^{iaA_\mu(\bfn)}\equiv -iU_\mu(\bfn), \nn \\
 A_\mu^\dagger(\bfn)&\to -\frac{1}{ia} e^{-iaA_\mu^\dagger(\bfn)} 
 \equiv i U_\mu^\dagger(\bfn) .
\label{shift}
\end{align}
To leading order in the dimensionful quantity $a$,
this is equivalent up to the usual shift prescription. In particular, in the
naive continuum limit we cannot tell the difference.
Note, however, that $U_\mu(\bfn)$ and $U_\mu^\dagger(\bfn)$ are not unitary matrices 
since $A_\mu(\bfn)$ and $A_\mu^\dagger(\bfn)$ are not 
hermitian. This point is crucial for what follows. For the moment, we can
choose to view the change $A_{\mu}(\bfn)  \to U_{\mu}(\bfn)$ as simply
a change of notation, since both $A_{\mu}(\bfn)$ and $U_{\mu}(\bfn)$
(although it notation-wise resembles a unitary link) are integrated over
as complex matrices.

As a result of these manipulations, we obtain a lattice action, 
\begin{align}
 S=\frac{1}{2g^2}\Tr Q_+\sum_{\bfn\in\Z_N^2}\Bigl\{
 &\chi_{\mu\nu}^\dagger(\bfn)\Bigl[
  -B_{\mu\nu}(\bfn)
  -i\Bigl(U_\mu(\bfn)U_\nu(\bfn+\bfe_\mu)
         -U_\nu(\bfn)U_\mu(\bfn+\bfe_\nu)\Bigr)\Bigr] \nn \\
 &+\chi_{\mu\nu}(\bfn)\Bigl[
  -B_{\mu\nu}^\dagger(\bfn)-i\Bigl(
  U_\mu^\dagger(\bfn+\bfe_\nu)U_\nu^\dagger(\bfn)
   -U_\nu^\dagger(\bfn+\bfe_\mu)U_\mu^\dagger(\bfn)\Bigr)
  \Bigr]
 \nn \\
 &- \psi_\mu^\dagger(\bfn) 
   \Bigl(U_\mu(\bfn)\bphi(\bfn+\bfe_\mu)-\bphi(\bfn)U_\mu(\bfn)\Bigr) \nn \\
 &- \psi_\mu(\bfn)
   \Bigl(U_\mu^\dagger(\bfn)\bphi(\bfn)-\bphi(\bfn+\bfe_\mu)U_\mu(\bfn)\Bigr) 
 \nn \\
 %&+\eta_-(\bfn)\Bigl(\frac{1}{8}d(\bfn)
 &+\frac{i}{4}\eta_+(\bfn)[\phi(\bfn),\bphi(\bfn)]
  +\frac{1}{2}\eta_-(\bfn)d(\bfn)
\Bigr\}, 
\label{Catterall's action}
\end{align}
where the BRST transformation (\ref{BRST charge pre 1}) becomes as 
\begin{alignat}{2}
 &Q_+ U_\mu(\bfn) = i\psi_\mu(\bfn), &\quad 
 &Q_+ \psi_\mu(\bfn) 
  = \frac{1}{2}\Bigl(
     U_\mu(\bfn)\phi(\bfn+\bfe_\mu)-\phi(\bfn)U_\mu(\bfn)\Bigr), \nn \\
 &Q_+ U_\mu^\dagger(\bfn) = -i\psi_\mu^\dagger(\bfn), &\quad 
  &Q_+ \psi_\mu^\dagger(\bfn) =  -\frac{1}{2}\Bigl(
     U_\mu^\dagger(\bfn)\phi(\bfn)-\phi(\bfn+\bfe_\mu)U_\mu(\bfn)\Bigr), \nn \\
 &Q_+ \bphi(\bfn) = \eta_+(\bfn), &\quad 
  &Q_+ \eta_+(\bfn) = -\frac{i}{2}[\phi(\bfn),\bphi(\bfn)], \nn \\
\label{BRST charge 1}  
 &Q_+ d(\bfn) = -\frac{1}{2}[\phi(\bfn),\eta_-(\bfn)], 
&\quad &Q_+ \eta_-(\bfn) = d(\bfn), \\
 &Q_+ \chi_{\mu\nu}(\bfn) = B_{\mu\nu}(\bfn), &\quad 
 &Q_+ B_{\mu\nu} = -\frac{i}{2}\Bigl(
  \phi(\bfn)\chi_{\mu\nu}(\bfn)
    -\chi_{\mu\nu}(\bfn)\phi(\bfn+\bfe_\mu+\bfe_\nu)\Bigr), \nn \\
 &Q_+ \chi_{\mu\nu}^\dagger(\bfn) = B_{\mu\nu}^\dagger(\bfn), &\quad 
 &Q_+ B_{\mu\nu}^\dagger(\bfn) = -\frac{i}{2}\Bigl(
  \phi(\bfn+\bfe_\mu+\bfe_\nu)\chi^\dagger_{\mu\nu}(\bfn)
    -\chi^\dagger_{\mu\nu}(\bfn)\phi(\bfn)
\Bigr), \nn \\
 &Q_+ \phi(\bfn)=0. \nn
\end{alignat}
Integrating out the auxiliary field $d(\bfn)$, 
the action (\ref{Catterall's action}) is nothing but that of 
the lattice gauge theory given in \cite{Catterall:2004np}. 
We emphasize that the prescription given in \cite{Catterall:2004np} 
is automatically reproduced by a combination of orbifolding and
the variant of deconstruction described above.

\subsection{Enhancement of symmetry by complexification}

The complexification of both bosonic and fermionic fields is reminiscent
of a balanced doubling of degrees of freedom on both the bosonic and
fermionic sides, and one is tempted to search for a
corresponding enhancement of supersymmetry. Indeed,
we can show that the complexified action (\ref{complex mother}) possesses 
another BRST-like symmetry, similar to the often encountered
additional anti-BRST symmetries of topological theories in the continuum. 
In fact, the action  can be rewritten as 
\begin{align}
 S=\frac{1}{2g^2}\Tr\Bigl\{ Q_+ Q_-\Bigl(
\frac{1}{2}\eta_- \eta_+ + 2\psi^\dagger_\mu \psi_\mu - \chi_{\mu\nu}^\dagger
 \chi_{\mu\nu}\Bigr)
+Q_+\Bigl(i\chi_{\mu\nu}^\dagger\left[A_\mu,A_\nu\right] 
        +i\chi_{\mu\nu}[A_\mu^\dagger,A_\nu^\dagger] \Bigr)
\Bigr\},
\label{complex action 2 SUSY}
\end{align}
where $Q_-$ acts on the fields as 
\begin{alignat}{2}
 &Q_- A_\mu = \psi_\mu, &\qquad &Q_- \psi_\mu = -\frac{i}{2}[A_\mu,\phi], \nn \\
 &Q_- A_\mu^\dagger = -\psi_\mu^\dagger, &\quad 
 &Q_- \psi_\mu^\dagger =  \frac{i}{2}[A_\mu^\dagger,\phi], \nn \\
 &Q_- \bphi = \eta_-, &\quad &Q_- \eta_+ = -d, \nn \\
\label{BRST charge pre 2}  
 &Q_- d = -\frac{i}{2}[\phi,\eta_+], 
  &\quad &Q_- \eta_- = \frac{i}{2}[\phi,\bphi], \\
 &Q_- \chi_{\mu\nu} = -B_{\mu\nu}, &\quad 
 &Q_- B_{\mu\nu} = -\frac{i}{2}[\phi,\chi_{\mu\nu}], \nn \\
 &Q_- \chi_{\mu\nu}^\dagger = B_{\mu\nu}^\dagger, &\quad 
 &Q_- B_{\mu\nu}^\dagger = \frac{i}{2}[\phi,\chi_{\mu\nu}^\dagger], 
 \qquad
 Q_- \phi =0, \nn 
\end{alignat}
and one can show that the second term of (\ref{complex action 2 SUSY}) 
is also $Q_-$-closed, $i.e.$ it is also manifestly $Q_-$-invariant. 
In fact, the second term can be expressed as 
\begin{equation}
 Q_-\Bigl(
  i\chi_{\mu\nu}^\dagger [A_\mu,A_\nu] 
 -i\chi_{\mu\nu}[A_\mu^\dagger,A_\nu^\dagger]
\Bigr). 
\end{equation}
$Q_-$ is also nilpotent up to gauge transformations and 
the two operators satisfy
\begin{equation}
 \left\{Q_+,Q_-\right\}=0 ~, 
\end{equation}
just like BRST and anti-BRST charges.

Note that the two supercharges $Q_+$ and $Q_-$ are actually 
independent of each
other, although the transformations 
(\ref{BRST charge pre 1}) and (\ref{BRST charge pre 2}) look quite similar. 
One way to see this is to use the relation between Catterall's complex
model and the orbifolded theory for two-dimensional $\cN=(4,4)$ supersymmetric
gauge theory \cite{Takimi:2007nn}. 
In the original orbifolded theory, there are two independent 
supercharges $Q$ and $\bar{Q}$, and they are not broken by the
truncation made in \cite{Takimi:2007nn}.
Using them, $Q_\pm$ can be written as $Q_\pm = (Q_+ \pm Q_-)/2$.

Correspondingly, the lattice action (\ref{Catterall's action}) can be
compactly written as 
\begin{align}
  S=\frac{1}{2g^2}\Tr \sum_{\bfn\in\Z_N^2}\biggl\{
 &Q_+Q_- \Bigl(
\frac{1}{2}\eta_-(\bfn) \eta_+(\bfn) 
+ 2\psi^\dagger_\mu(\bfn) \psi_\mu(\bfn) 
- \chi_{\mu\nu}^\dagger(\bfn) \chi_{\mu\nu}(\bfn)\Bigr) \nn \\
&+Q_+\Bigl(
\chi_{\mu\nu}^\dagger(\bfn)\Bigl[
  -B_{\mu\nu}(\bfn)
  -i\Bigl(U_\mu(\bfn)U_\nu(\bfn+\bfe_\mu)
         -U_\nu(\bfn)U_\mu(\bfn+\bfe_\nu)\Bigr)\Bigr] \nn \\
 &\hspace{1cm}+\chi_{\mu\nu}(\bfn)\Bigl[
  -B_{\mu\nu}^\dagger(\bfn)-i\Bigl(
  U_\mu^\dagger(\bfn+\bfe_\nu)U_\nu^\dagger(\bfn)
   -U_\nu^\dagger(\bfn+\bfe_\mu)U_\mu^\dagger(\bfn)\Bigr)
  \Bigr]\Bigr)\biggr\}, 
\end{align}
where the BRST charge $Q_-$ acts in the following manner: 
\begin{alignat}{2}
 &Q_- U_\mu(\bfn) = i\psi_\mu(\bfn), &\quad 
 &Q_- \psi_\mu(\bfn) 
  = -\frac{1}{2}\Bigl(
     U_\mu(\bfn)\phi(\bfn+\bfe_\mu)-\phi(\bfn)U_\mu(\bfn)\Bigr), \nn \\
 &Q_- U_\mu^\dagger(\bfn) = i\psi_\mu^\dagger(\bfn), &\quad 
  &Q_- \psi_\mu^\dagger(\bfn) =  -\frac{1}{2}\Bigl(
     U_\mu^\dagger(\bfn)\phi(\bfn)-\phi(\bfn+\bfe_\mu)U_\mu(\bfn)\Bigr), \nn \\
\label{BRST charge 2}  
 &Q_- \bphi(\bfn) = \eta_-(\bfn), &\quad 
  &Q_- \eta_+(\bfn) = -d(\bfn), \\
 &Q_- d(\bfn) = -\frac{1}{2}[\phi(\bfn),\eta_+(\bfn)], 
&\quad &Q_- \eta_-(\bfn) = \frac{i}{2}[\phi(\bfn),\bphi(\bfn)], \nn \\
 &Q_- \chi_{\mu\nu}(\bfn) = -B_{\mu\nu}(\bfn), &\quad 
 &Q_- B_{\mu\nu}(\bfn) = -\frac{i}{2}\Bigl(
  \phi(\bfn)\chi_{\mu\nu}(\bfn)
    -\chi_{\mu\nu}(\bfn)\phi(\bfn+\bfe_\mu+\bfe_\nu)\Bigr), \nn \\
 &Q_- \chi_{\mu\nu}^\dagger(\bfn) = B_{\mu\nu}^\dagger(\bfn), &\quad 
 &Q_- B_{\mu\nu}^\dagger(\bfn) = \frac{i}{2}\Bigl(
  \phi(\bfn+\bfe_\mu+\bfe_\nu)\chi^\dagger_{\mu\nu}(\bfn)
    -\chi^\dagger_{\mu\nu}(\bfn)\phi(\bfn)
\Bigr), \nn \\
 &Q_- \phi(\bfn)=0. \nn
\end{alignat}

\subsection{Naive reduction back to the real line}

Because complexification played such a crucial role in deriving the
supersymmetric lattice action (\ref{Catterall's action}), we should
expect difficulties if we {\em a posteriori} reduce fields from
the complex plane back to the real line. Indeed, there are problems
at many different levels. Let us first consider the lattice gauge
symmetry of the complexified action. From the orbifolding procedure
the ultralocal $U(k)$ symmetry of the zero-dimensional mother
theory becomes a lattice gauge symmetry, where fields transform
as either adjoints or bifundamentals, $viz.$,
\begin{alignat}{2}
U_{\mu}(\bfn) &\to V^{\dagger}(\bfn)U_{\mu}(\bfn)V(\bfn+\bfe_\mu), &\quad 
U^{\dagger}_{\mu}(\bfn) &\to
 V(\bfn+\bfe_\mu)U^{\dagger}_{\mu}(\bfn)V^{\dagger}(\bfn), \nn \\
\phi(\bfn) &\to V^{\dagger}(\bfn)\phi(\bfn)V(\bfn), &\quad 
\bphi(\bfn) &\to
 V(\bfn)\bphi(\bfn)V^{\dagger}(\bfn), \nn \\
B_{12}(\bfn) &\to V^{\dagger}(\bfn)B_{12}(\bfn)V(\bfn+\bfe_1+\bfe_2), &\quad 
B_{12}^{\dagger}(\bfn) &\to
 V(\bfn+\bfe_1+\bfe_2)B^{\dagger}_{12}(\bfn)V^{\dagger}(\bfn), \nn \\
\label{gaugetransf}
\psi_{\mu}(\bfn) &\to V^{\dagger}(\bfn)\psi_{\mu}(\bfn)V(\bfn+\bfe\mu), &\quad 
\psi^{\dagger}_{\mu}(\bfn) &\to
 V(\bfn+\bfe_\mu)\psi^{\dagger}_{\mu}(\bfn)V^{\dagger}(\bfn),  \\
\eta_\pm(\bfn) &\to V^{\dagger}(\bfn)\eta_\pm(\bfn)V(\bfn), &\quad 
d(\bfn) &\to
 V(\bfn)d(\bfn)V^{\dagger}(\bfn), \nn \\
\chi_{12}(\bfn) &\to
 V^{\dagger}(\bfn)\chi_{12}(\bfn)V(\bfn+\bfe_1+\bfe_2), &\quad 
\chi_{12}^{\dagger}(\bfn) &\to
 V(\bfn+\bfe_1+\bfe_2)\chi^{\dagger}_{12}(\bfn)V^{\dagger}(\bfn), \nn
\end{alignat}
where $V \in U(k)$.
In a first attempt at projecting onto the real axis, one could consider
\cite{Catterall:2004np} 
taking $A_{\mu}(\bfn)$ hermitian, and hence $U_{\mu}(\bfn)$ unitary.
This does not alter the gauge transformation for $U_{\mu}$. But reducing the
other fields from being complex to being hermitian is not compatible with
the $U(k)$ symmetry. For instance, requiring $\psi_{\mu}(\bfn)
=\psi^{\dagger}_{\mu}(\bfn)$ is clearly incompatible with the general
gauge transformation rule (\ref{gaugetransf}).

Another difficulty with a naive reduction to the real line is the
breaking of the BRST--anti-BRST symmetries. Clearly, if we take $A_{\mu}(\bfn)$
to be hermitian, and thus $U_{\mu}(\bfn)$ unitary, the supersymmetry
transformations $Q_{\pm}U_{\mu}(\bfn) = i\psi_{\mu}(\bfn)$ and
$Q_{\pm}U^{\dagger}_{\mu}(\bfn) = \mp i\psi^{\dagger}_{\mu}(\bfn)$ are
incompatible with the unitarity constraint $U_{\mu}(\bfn)U^{\dagger}_{\mu}(\bfn)
= 1$. One consequence of this incompatibility is a breaking of the
remnants of supersymmetry already at the action level. This is as 
expected, since one must impose the unitarity constraint
$U_{\mu}(\bfn)U^{\dagger}_{\mu}(\bfn) = 1$ in the action, while
one needs $Q_{\pm}\left(U_{\mu}(\bfn)U^{\dagger}_{\mu}(\bfn)\right) \neq 0$ in
order for the action to remain invariant under $Q_{\pm}$. One can check
explicitly that this breaking of supersymmetry occurs in the action.

Related to this is the incompatibility of the supersymmetry transformations
$Q_{\pm}U_{\mu}(\bfn) = i\psi_{\mu}(\bfn)$ with invariances of the
functional measure. In the continuum, topological field theories are
based on the largest invariance possible,
\begin{equation}
QA_{\mu}(x) ~=~ \psi_{\mu}(x) ~, \label{contshift}
\end{equation}
of the gauge potential $A_{\mu}(x)$. This corresponds to the most general
shift symmetry of the measure in that case. For the unitary lattice
variable $U_{\mu}(x)$, which should be integrated over the left and
right invariant Haar measure, there is no corresponding shift symmetry.
Instead, the analog of general shift symmetry corresponds to the most
general motion on the unitary group manifold. This is not generated by
an ordinary derivative, but by the Lie derivative $\nabla^a$. Infinitesimally,
this requires a supersymmetry transformation rule for $U_{\mu}(\bfn)$
of, for a left derivative, 
\begin{equation}
QU_{\mu}(\bfn) = i\psi_{\mu}(\bfn)U_{\mu}(\bfn) ~,
\label{Sugino QU}
\end{equation}
and this is indeed the direct lattice analog of the continuum 
transformation (\ref{contshift}). The Haar measure is invariant under
such a transformation, and it is of course also
by construction compatible with the unitarity constraint 
$U_{\mu}(\bfn)U^{\dagger}_{\mu}(\bfn) = 1$. The Haar measure is {\em
not} invariant under the naive rule $Q U_\mu = i \psi_\mu$, 
with $U_{\mu}$ unitary.
Supersymmetry is therefore broken in both the action and the
measure (and the combination of the two). 

Remarkably, lattice Monte Carlo simulations \cite{Catterall:2006jw} 
indicate
that the actual breaking of supersymmetry with this kind of reduction
to the real line, even at quite strong
coupling, is almost undetectable. Perhaps the reason is that the
degrees of freedom are correctly specified in terms of the ``natural''
fermionic variables (site variables, link variables, and corner
variables), and that the number of bosonic and fermionic degrees
match. This issue deserves more attention, as it may point towards
new and approximate manners of simulating supersymmetric field
theories on the lattice. 

\section{Comment on a relation to Sugino's lattice action}

Very recently, Takimi \cite{Takimi:2007nn} 
has shown how a small deformation of Sugino's lattice formulation 
of two-dimensional $\cN=(2,2)$ supersymmetric gauge theory 
\cite{Sugino:2003yb}\cite{Sugino:2004qd} can be obtained
by a consistent truncation of some of the degrees of freedom in 
Catterall's model, while still preserving a BRST symmetry. 
In this section, we make some comments on this truncation. In particular,
since we have now realized that there are in fact two scalar supersymmetries,
we wish to see what happens if we instead perform a similar truncation
that preserves the other (anti-)BRST charge.

Let us first briefly review the idea of ref.~\cite{Takimi:2007nn}. 
First of all, we regard $U_\mu(\bfn)$ as unitary matrices so
that $U_\mu(\bfn)U_\mu^\dagger(\bfn)=1$. 
By this truncation, we impose hermiticity of $A_\mu(\bfn)$.
In order that this truncation is consistent with the BRST transformation 
by $Q_+$, we impose 
\begin{equation}
 Q_+ (U_\mu(\bfn)U_\mu^\dagger(\bfn))=0,
\end{equation}
which leads to
\begin{equation}
 \psi_\mu^\dagger(\bfn)
 = U_\mu^\dagger(\bfn) \psi_\mu(\bfn) U_\mu^\dagger(\bfn), 
 \label{psi dagger}
\end{equation}
or equivalently, 
\begin{equation}
 \left(\psi_{(\mu)}(\bfn)\right)^\dagger = \psi_{(\mu)}, \qquad 
 \psi_{(\mu)}(\bfn) \equiv \psi_\mu(\bfn)U_\mu^\dagger(\bfn),
\end{equation}
that is, $\psi_{(\mu)}(\bfn)$ are hermitian. 
Here, the link variables $\psi_\mu(\bfn)$ have been 
transformed into {\em site variables} $\psi_{(\mu)}(\bfn)$. 
Similarly, we define a site variable, 
\begin{align}
 \chi(\bfn)&\equiv \chi_{12}(\bfn)U_2^\dagger(\bfn+\bfe_1) 
 U_1^\dagger(\bfn), \nn \\
\end{align}
and impose it to be hermitian.
Then, $\chi_{12}^\dagger(\bfn)$ is related to $\chi_{12}(\bfn)$ as 
\begin{equation}
 \chi_{12}^\dagger(\bfn)=U_2^\dagger(\bfn+\bfe_1)U_1^\dagger(\bfn)
  \chi_{12}(\bfn)U_2^\dagger(\bfn+\bfe_1)U_1^\dagger(\bfn).
\end{equation}
Furthermore, we define a hermitian field $H(\bfn)$ through the relation, 
\begin{equation}
 B_{12}(\bfn)=H(\bfn)U_1(\bfn)U_2(\bfn+\bfe_1)
 -i\chi(\bfn)\Bigl(\psi_1(\bfn)U_2(\bfn+\bfe_1)
  +U_1(\bfn)\psi_2(\bfn+\bfe_1)\Bigr). 
\end{equation}
As same as the case of $\chi_{12}^\dagger$, 
$B_{12}^\dagger$ is determined uniquely by imposing $H(\bfn)$ to be
hermitian:
\begin{equation}
  B_{12}^\dagger(\bfn)=U_2^\dagger(\bfn+\bfe_1)U_1^\dagger(\bfn)H(\bfn)
 -i
  \Bigl(U_2^\dagger(\bfn+\bfe_1)\psi_1^\dagger(\bfn)
  +\psi_2^\dagger(\bfn+\bfe_1)U_1^\dagger(\bfn)
  \Bigr)\chi(\bfn). 
\end{equation}
Finally, we set 
\begin{equation}
 \eta_+(\bfn)\equiv \eta(\bfn), \quad \eta_-(\bfn)\equiv 0, \quad 
 d(\bfn)\equiv 0. 
\end{equation}
As a result, the BRST transformation (\ref{BRST charge 1}) turns out to
be 
\begin{alignat}{2}
 &Q U_\mu(\bfn)= i\psi_{(\mu)}(\bfn)U_\mu(\bfn), &\quad 
  & Q \psi_{(\mu)}(\bfn)=i \psi_{(\mu)}(\bfn)\psi_{(\mu)}(\bfn)
   +\frac{1}{2}\Bigl(U_\mu(\bfn)\phi(\bfn+\bfe_\mu)U_\mu^\dagger(\bfn)
     -\phi(\bfn)\Bigr), \nn \\
 &Q  \bphi(\bfn)=\eta(\bfn), &\quad 
&Q \eta(\bfn) = -\frac{i}{2}[\phi(\bfn),\bphi(\bfn)], \nn \\
 &Q \chi(\bfn)= H(\bfn), &\quad 
&Q H(\bfn)=-\frac{1}{2}[\phi(\bfn),\chi(\bfn)], 
\label{BRST Sugino}
\end{alignat}
with $Q \equiv Q_+$. 
This is nothing but the BRST transformation of Sugino's lattice
formulation of the two-dimensional $\cN=(2,2)$ supersymmetric gauge
theory \cite{Sugino:2004qd} and the consistent BRST transformation 
(\ref{Sugino QU}) has been automatically derived. 
One can also show that the action of Catterall's model 
(\ref{Catterall's action}) turns out to be almost that of Sugino's model 
by this truncation of degrees of freedom (for details, see
\cite{Takimi:2007nn}).

An immediate question is whether the anti-BRST symmetry $Q_-$ 
is preserved or not. One can easily see that this is not the case. 
In fact, under the truncation adopted above, the anti-BRST transformation 
of $U_\mu(\bfn)U_\mu(\bfn)^\dagger$ (the combination that equals
unity if $U_\mu$ is restricted to be unitary) 
under the action of $Q_-$ is not zero:%
%\footnote{The same discussion can be applied to 
%Catterall's original prescription in which one restricts the path
%integral to the ``real line'', $\psi_\mu=\psi_\mu^\dagger$.}
\begin{equation}
 Q_-(U_\mu(\bfn)U_\mu(\bfn)^\dagger) = 2i\psi_{(\mu)}(\bfn) \ne 0 ~. 
\end{equation}
%that is, the transformation by $Q_-$ makes $U_\mu(\bfn)$ to 
%non-unitary matrices in general. 
Similarly, we can show that the action of $Q_-$ 
is incompatible with hermiticity of 
$\chi(\bfn)$ and $H(\bfn)$ and the conditions $d(\bfn) = \eta_-(\bfn) = 0$. 
Therefore, $Q_-$ is not consistent with the rule of truncation introduced
above, and the truncated theory possesses only one preserved BRST
charge.

In the above argument, we truncated the degrees of freedom with
preserving the BRST symmetry $Q_+$. 
However, in principle, we can choose any linear combination of $Q_+$
and $Q_-$ to be preserved.  
As example, let us choose $Q_-$ to be preserved. 
In this case, the relation corresponding to (\ref{psi dagger}) is 
\begin{equation}
 \psi_\mu^\dagger(\bfn)=-U_\mu^\dagger(\bfn)\psi_\mu(\bfn)U_\mu^\dagger(\bfn),
\end{equation}
then we can define hermitian site fermions as, 
\begin{equation}
 \psi_{(\mu)}(\bfn) = i\psi_\mu(\bfn)U_\mu^\dagger(\bfn).
\end{equation} 
Similarly, we can define hermitian site variables $\chi(\bfn)$
and $H(\bfn)$ by 
\begin{align}
 \chi(\bfn)&=\chi_{12}(\bfn)U_2^\dagger(\bfn+\bfe_1)U_1^\dagger(\bfn), \\
 H(\bfn)&= i B_{12}(\bfn)U_2^\dagger(\bfn+\bfe_1)U_1^\dagger(\bfn)
  -i\chi(\bfn)\psi_{(2)}(\bfn+\bfe_1)U_1^\dagger(\bfn)
  -i\chi(\bfn)\psi_{(1)}(\bfn), 
\end{align}
which is consistent with the BRST transformation (\ref{BRST charge 2}). 
We can also restrict $\eta_\pm(\bfn)$ and $d(n)$ as 
\begin{equation}
 \eta_+(\bfn)\equiv0, \quad \eta_-(\bfn) \equiv i\eta(\bfn), \quad 
  d(\bfn)\equiv 0. 
\end{equation}
By this truncation, we obtain the same BRST transformation as
(\ref{BRST Sugino}) after setting $Q\equiv -iQ_-$ and we again obtain 
the action of Sugino's formulation (plus the additional terms). 
In this case, the BRST symmetry $Q_+$ is broken after the truncation. 
The argument is completely parallel for any linear combination of 
$Q_+$ and $Q_-$, 
\begin{equation}
 \tilde{Q} \equiv \alpha Q_+ + \beta Q_- ~. 
\end{equation}
If $\beta=\pm \alpha$ 
it seems impossible to impose 
the condition $U_\mu^\dagger(\bfn)U_\mu(\bfn)=1$.

\section{Application to Four-dimensional $\cN=2$ Supersymmetric
 Yang-Mills Theory}

As mentioned in the introduction, we can apply the prescription
discussed in the section 2 to any other supersymmetric gauge theory. 
In particular, it seems to be also applicable to such a theory that is
not described by Dirac-K\"{a}hler fermions. 
In this section, we apply it to four-dimensional $\cN=2$
supersymmetric Yang-Mills theory as an example. 

The starting point of the discussion is the mother theory, that is, 
the dimensionally reduced theory of the four-dimensional $\cN=2$
supersymmetric Yang-Mills Lagrangian. 
The purpose is to construct a lattice formulation that possesses at
least one supercharge. 
To this end, we start with the dimensional reduced action of 
the topologically twisted
four-dimensional $\cN=2$ SYM theory
\cite{Witten:1988ze}: 
\begin{align}
S=\frac{1}{g^2}\Tr Q \Bigl\{
-\chi_{\mu\nu}^+ \Bigl(B_{\mu\nu}^+-F_{\mu\nu}\Bigr) 
-\frac{i}{2}\psi_\mu[A_\mu,\bphi]
+\frac{i}{8}\eta[\phi,\bphi]
\Bigr\}, 
\label{N=2 mother}
\end{align}
where $\mu,\nu=1,\cdots,4$ and $F_{\mu\nu}\equiv i[A_\mu,A_\nu]$. 
We have assumed that 
$\{A_\mu, \bphi, B_{\mu\nu}^+, \phi\}$ 
and $\{\psi_\mu, \eta, \chi_{\mu\nu}^+\}$ 
are bosonic and fermionic hermitian matrices with
the size $kN^4$, respectively, and 
$\chi_{\mu\nu}^+$ and $B_{\mu\nu}^+$ are anti-symmetric with respect to
the Lorentz indices and satisfy the self-dual condition, 
$\frac{1}{2}\e_{\mu\nu\rho\sigma}\chi_{\rho\sigma}^+=\chi_{\mu\nu}^+$ 
and the same equation for $B_{\mu\nu}^+$. 
The BRST charge $Q$ acts on the fields as 
\begin{alignat}{2}
 &Q A_\mu = \psi_\mu, &\qquad &Q \psi_\mu = -i[A_\mu,\phi],  \nn \\
\label{N=2 BRST trans}
 &Q \bphi = \eta, \quad &\quad &Q \eta = i[\phi,\bphi], \\
 &Q \chi_{\mu\nu}^+ = B_{\mu\nu}^+, &\quad 
  &Q B_{\mu\nu}^+ = i[\phi,\chi_{\mu\nu}^+], \qquad
 \quad Q\phi =0.  \nn
\end{alignat}

As we did in the section 2, 
we next extend the theory by complexifying 
the fields $A_\mu$, $\psi_\mu$, $\chi_{\mu\nu}^+$ 
and $B_{\mu\nu}^+$ in order that the theory has enough $U(1)$ symmetries
to create four-dimensional space-time by orbifolding. 
In this case, however, the complexification is not sufficient, 
since the self-duality of the fields $\chi_{\mu\nu}^+$ and
$B_{\mu\nu}^+$ makes it impossible to define $U(1)$ charges that is
compatible with the first term of the action (\ref{N=2 mother}). 
To overcome this problem, we further extend $\chi_{\mu\nu}^+$ and
$B_{\mu\nu}^+$ to complex rank 2 tensors without self-dual constraint, 
$\chi_{\mu\nu}$ and $B_{\mu\nu}$, respectively. 
After these extension, we obtain the action of 
``complexified'' mother theory: 
\begin{align}
 S=\frac{1}{2g^2}\Tr Q\Bigl\{
 -\chi_{\mu\nu}^\dagger &\Bigl(B_{\mu\nu}-F_{\mu\nu}\Bigr)
 -\chi_{\mu\nu}\Bigl(B_{\mu\nu}^\dagger-F_{\mu\nu}^\dagger\Bigr) \nn \\
 &-\frac{i}{2}\psi_\mu^\dagger [A_\mu,\bphi]
 -\frac{i}{2}\psi_\mu [A_\mu^\dagger,\bphi]
+\frac{i}{4}\eta[\phi,\bphi]\Bigr\}. 
\label{N=2 cpx mother pre}
\end{align}

\begin{table}[h]
\caption{The charge assignment for the complexified fields} 
\begin{center}
\begin{tabular}{c|ccccccccccc}
 & $A_\mu$ & $A_\mu^\dagger$ & $\phi$ & $\bphi$ & $B_{\mu\nu}$ 
 & $B_{\mu\nu}^\dagger$ & $\eta$ & $\psi_\mu$ & $\psi_\mu^\dagger$ 
 & $\chi_{\mu\nu}$ & $\chi_{\mu\nu}^\dagger$ \\
\hline
${\mathbf q}$ & $\bfe_\mu$ & $-\bfe_\mu$ & 0 & 0 & $\bfe_\mu+\bfe_\nu$ 
& $-\bfe_\mu-\bfe_\nu$ & 0 & $\bfe_\mu$ & $-\bfe_\mu$ 
& $\bfe_\mu+\bfe_\nu$ &  $-\bfe_\mu-\bfe_\nu$ \\
\end{tabular}
\end{center}
\label{N=2 charge table}
\end{table}

For the fields in this complexified theory, we can assign non-trivial
$U(1)$ charges as in Table \ref{N=2 charge table}, 
where ${\mathbf q}\equiv(q_1,q_2,q_3,q_4)$ is a set 
of four $U(1)$ charges 
and we have defined 
\begin{equation}
 \bfe_1 \equiv \left(\begin{matrix}1 \\ 0 \\ 0 \\ 0 \end{matrix}\right),
 \quad 
 \bfe_2 \equiv \left(\begin{matrix}0 \\ 1 \\ 0 \\ 0 \end{matrix}\right),
 \quad 
 \bfe_3 \equiv \left(\begin{matrix}0 \\ 0 \\ 1 \\ 0 \end{matrix}\right),
 \quad 
 \bfe_4 \equiv \left(\begin{matrix}0 \\ 0 \\ 0 \\ 1
                     \end{matrix}\right). 
\end{equation}
Correspondingly, we can make orbifolding by substituting the
expansion like (\ref{orbifolding}) into the complexified action
(\ref{N=2 cpx mother pre}). 
The lattice action obtained by carrying out the replacement like
(\ref{shift}) followed by some consistent truncation of the degrees of
freedom.

In order to simplify the description, however, we change the order of
the prescription in this section; 
(1) we first replace $A_\mu$ to $-iU_\mu$ (deconstruction), 
(2) we next truncate some degrees of freedom of the complexified 
matrix theory, 
and (3) we finally will perform the orbifolding. 
We can explicitly show that it is equivalent to the prescription
discussed in the section 2.

Following this procedure, we first replace $A_\mu$ and $A_\mu^\dagger$ by 
\begin{equation}
 A_\mu \to -i U_\mu, \quad A_\mu^\dagger \to i U_\mu^\dagger. 
\end{equation}
Then the action (\ref{N=2 cpx mother pre}) becomes 
\begin{align}
 S=\frac{1}{2g^2}\Tr Q\Bigl\{
 -\chi_{\mu\nu}^\dagger &\Bigl(B_{\mu\nu}-\cF_{\mu\nu}\Bigr)
 -\chi_{\mu\nu}\Bigl(B_{\mu\nu}^\dagger-\cF_{\mu\nu}^\dagger\Bigr) \nn \\
&-\frac{1}{2}\psi_\mu^\dagger [U_\mu,\bphi]
 +\frac{1}{2}\psi_\mu [U_\mu^\dagger,\bphi]
+\frac{i}{4}\eta[\phi,\bphi]\Bigr\},  
\label{N=2 cpx mother}
\end{align}
where $\cF_{\mu\nu}$ is given by
\begin{equation}
\cF_{\mu\nu}=-i[U_\mu,U_\nu],
\end{equation}
and the BRST transformation (\ref{N=2 BRST trans}) becomes 
\begin{alignat}{2}
 &Q U_\mu = i\psi_\mu, &\qquad &Q \psi_\mu = -[U_\mu,\phi],  \nn \\
 &Q U_\mu^\dagger = -i\psi_\mu^\dagger, &\quad 
  &Q \psi_\mu^\dagger = [U_\mu^\dagger,\phi],  \nn \\
\label{N=2 cpx BRST trans}
 &Q \bphi = \eta, &\quad &Q \eta = i[\phi,\bphi], \\
 &Q \chi_{\mu\nu} = B_{\mu\nu}, &\quad 
  &Q B_{\mu\nu} = i[\phi,\chi_{\mu\nu}], \nn \\ 
 &Q \chi_{\mu\nu}^\dagger = B_{\mu\nu}^\dagger, &\quad 
  &Q B_{\mu\nu}^\dagger = i[\phi,\chi_{\mu\nu}^\dagger], \qquad 
 \quad Q\phi =0.  \nn
\end{alignat}

Next, we must truncate some degrees of freedom. 
As discussed in the section 2.3, the naive restriction 
to ``real line'' breaks not only the remaining supersymmetry 
but also the gauge symmetry of the system, in general. 
Thus, it seems to be better to adopt the way of truncation 
adopted in \cite{Takimi:2007nn}. 
We first impose $U_\mu$ to be unitary matrices. 
Then, repeating the same discussion around (\ref{psi dagger}), 
we can show that $\psi_\mu^\dagger$ is related to $\psi_\mu$ as 
\begin{equation}
 \psi_\mu^\dagger = U_\mu^\dagger \psi_\mu U_\mu^\dagger, 
\label{N=2 psi dagger}
\end{equation}
and we can define hermitian matrices $\psi_{(\mu)}$ as 
\begin{equation}
 \psi_{(\mu)} \equiv \psi_\mu U_\mu^\dagger. 
 \label{N=2 hermitian psi}
\end{equation}
In (\ref{N=2 psi dagger}) and (\ref{N=2 hermitian psi}), we do {\em not} 
sum over $\mu$. 
In the following, we do not sum over duplicated symbols unless we
explicitly write it. 

In order to truncate the half of the degrees of freedom of $\chi_{\mu\nu}$
we define complex fermionic 
fields $\chi_{(\mu\nu)}$ with zero $U(1)$ charges as 
\begin{equation}
 \chi_{(\mu\nu)} = 
\begin{cases}
\chi_{\mu\nu} U_\nu^\dagger U_\mu^\dagger, \quad 
 &{\rm for}\ \ (\mu,\nu)\in {\cI} \\
 -\chi_{\nu\mu} U_\mu^\dagger U_\nu^\dagger, \quad 
 &{\rm for}\ \ (\mu,\nu) \in\!\!\!\!\!\slash~ \cI
\end{cases} 
\label{scalar chi}
\end{equation}
where 
\begin{equation}
 \cI \equiv \{(1,4),(2,4),(3,4),(2,3),(3,1),(1,2)\}, 
\end{equation}
and impose $\chi_{(\mu\nu)}$ to be hermitian. 
The new field $\chi_{(\mu\nu)}$ satisfies 
$\chi_{(\mu\nu)}=-\chi_{(\nu\mu)}$ by definition. 
Note that we can impose the hermiticity only for those fields which have
zero $U(1)$ charges. 
Correspondingly, we define bosonic hermitian 
anti-symmetric tensor field
$H_{(\mu\nu)}$ through the BRST transformation: 
\begin{equation}
 Q \chi_{(\mu\nu)} \equiv H_{(\mu\nu)}. 
\end{equation}
The original fields $\chi_{\mu\nu}$ and $B_{\mu\nu}$ can be expressed by 
the new fields as 
\begin{align}
 \psi_\mu &= \psi_{(\mu)}U_{\mu}, \nn \\
 \chi_{\mu\nu} &= \chi_{(\mu\nu)} U_{\mu}U_{\nu}, 
\qquad ({\rm for}\ (\mu,\nu)\in \cI)  \\
 B_{\mu\nu} &= H_{(\mu\nu)}U_{\mu}U_\nu -i \chi_{(\mu\nu)}\Bigl(
 U_\mu \psi_{(\mu)}U_\nu + \psi_{(\mu)}U_{\mu}U_\nu
\Bigr).\nn 
\end{align}
We must further restrict the degrees of freedom of $\chi_{(\mu\nu)}$ 
and $H_{(\mu\nu)}$, and it seems to be proper to impose the self-dual 
condition to them: 
\begin{equation}
 \frac{1}{2}\sum_{\rho,\sigma=1}^4
\e_{\mu\nu\rho\sigma}\chi_{(\rho\sigma)} = \chi_{(\mu\nu)}, 
\quad 
 \frac{1}{2}\sum_{\rho,\sigma=1}^4
\e_{\mu\nu\rho\sigma}H_{(\rho\sigma)} = H_{(\mu\nu)}. 
\label{self-dual again}
\end{equation}
{}From now on, we denote the three independent components of
$\chi_{(\mu\nu)}$ and $H_{(\mu\nu)}$ as 
\begin{align}
 \vec{\chi} &\equiv \left(\chi_1,\chi_2,\chi_3\right) 
           \equiv \left({2}\chi_{(14)},{2}\chi_{(24)},{2}\chi_{(34)}\right),
 \nn \\
 \vec{H}  &\equiv \left(H_1,H_2,H_3\right) 
           \equiv \left({2}H_{(14)},2H_{(24)},2H_{(34)}\right),
\end{align}
After the above truncation, the action (\ref{N=2 cpx mother}) becomes 
\begin{align}
 S=\frac{1}{g^2}\Tr Q\Bigl\{
-\vchi\cdot\left(\vH+\vPhi \right)
+\frac{1}{2}\sum_{\mu=1}^4
\psi_{(\mu)}\left(\bphi-U_\mu\bphi U_\mu^\dagger \right)
+\frac{i}{8}\eta[\phi,\bphi]
+\frac{i}{2}\sum_{i=1}^3 \chi_i\Psi_i\chi_i
\Bigr\},
\label{truncated matrix action}
\end{align}
where $\vPhi=(\Phi_1,\Phi_2,\Phi_3)$ is given by 
\begin{align}
 \Phi_1 &= \frac{i}{2}\left(U_{14}-U_{41}+U_{23}-U_{32}\right), \nn \\
 \Phi_2 &= \frac{i}{2}\left(U_{24}-U_{42}+U_{31}-U_{13}\right), \\
 \Phi_3 &= \frac{i}{2}\left(U_{34}-U_{43}+U_{12}-U_{21}\right), \nn 
\end{align} 
with 
\begin{equation}
U_{\mu\nu} \equiv U_{\mu}U_\nu U_\mu^\dagger U_\nu^\dagger,  
\end{equation}
and $\vPsi=(\Psi_1,\Psi_2,\Psi_3)$ is given by 
\begin{align}
 \Psi_1 &= \cL^+_{4}\psi_{(1)}+\cL^+_{1}\psi_{(4)} 
          +\cL^+_{3}\psi_{(2)}+\cL^+_{2}\psi_{(3)}, \nn \\
 \Psi_2 &= \cL^+_{4}\psi_{(2)}+\cL^+_{2}\psi_{(4)} 
          +\cL^+_{1}\psi_{(3)}+\cL^+_{3}\psi_{(1)}, \\
 \Psi_3 &= \cL^+_{4}\psi_{(3)}+\cL^+_{3}\psi_{(4)} 
          +\cL^+_{2}\psi_{(1)}+\cL^+_{1}\psi_{(2)}, \nn 
\end{align}
where 
\begin{equation}
 \cL^+_{\nu}\psi_{(\mu)}\equiv \psi_{(\mu)}+U_\nu \psi_{(\mu)} U_\nu^\dagger.
\end{equation}
The BRST transformation (\ref{N=2 cpx BRST trans}) becomes 
\begin{alignat}{2}
 &Q U_\mu = i\psi_{(\mu)} U_\mu, &\qquad  
 &Q \psi_{(\mu)} = \phi - U_\mu \phi U_\mu^\dagger 
  +i \psi_{(\mu)}\psi_{(\mu)} \nn \\
\label{N=2 truncated BRST}
 &Q\bphi = \eta, & &Q\eta = i[\phi,\bphi], \\
 &Q\vchi = \vH, & &Q\vH = i[\phi,\vchi], 
\qquad Q\phi =0. \nn 
\end{alignat}

Finally, we generate a lattice action from 
the truncated action (\ref{truncated matrix action}) by orbifolding. 
By construction, the $U(1)$ charges of $U_\mu$ are given by
$\bfe_\mu$ and those of other fields are zero. 
Then, the orbifold projection is achieved by substituting 
the following expansions into the truncated action 
(\ref{truncated matrix action}): 
\begin{alignat}{2}
 &U_\mu = \sum_{\bfn\in\Z_N^4} U_\mu(\bfn) \otimes E_{\bfn,\bfn+\bfe_\mu}, 
 &\qquad 
 &U_\mu^\dagger = \sum_{\bfn\in\Z_N^4} U_\mu^\dagger(\bfn)
 \otimes E_{\bfn+\bfe_\mu,\bfn}, \nn \\
 &\phi = \sum_{\bfn\in\Z_N^4} \phi(\bfn) \otimes E_{\bfn,\bfn},
 &\qquad 
 &\bphi = \sum_{\bfn\in\Z_N^4} \bphi(\bfn) \otimes E_{\bfn,\bfn}, \nn \\
 &\psi_{(\mu)} =  \sum_{\bfn\in\Z_N^4} 
  \psi_{(\mu)}(\bfn) \otimes E_{\bfn,\bfn}, 
 &\quad 
 &\eta = \sum_{\bfn\in\Z_N^4} \eta(\bfn) \otimes E_{\bfn,\bfn},  \\
 &\vchi =  \sum_{\bfn\in\Z_N^4} \vchi(\bfn) \otimes E_{\bfn,\bfn}, 
 &\quad 
 &\vH = \sum_{\bfn\in\Z_N^4} \vH(\bfn) \otimes E_{\bfn,\bfn}, \nn
\end{alignat}
where link variables $U_\mu(\bfn)$ take values in $U(k)$ and 
the other lattice fields are hermitian matrices with the size $k$. 
As a result, we obtain the action of a lattice formulation 
for the topologically twisted 
four-dimensional $\cN=2$ supersymmetric Yang-Mills theory: 
\begin{align}
 S=\frac{1}{g^2}\Tr\sum_{\bfn\in\Z_N^4} Q\Bigl\{
-\vchi(\bfn)\cdot&\left(\vH(\bfn)+\vPhi(\bfn) \right)
+\frac{1}{2}\sum_{\mu=1}^4 \psi_{(\mu)}(\bfn)\left(\bphi(\bfn)
-U_\mu(\bfn)\bphi(\bfn+\bfe_\mu) U_\mu^\dagger(\bfn) \right) \nn \\
&+\frac{i}{8}\eta(\bfn)[\phi(\bfn),\bphi(\bfn)]
+\frac{i}{2}\sum_{i=1}^3 \chi_i(\bfn)\Psi_i(\bfn)\chi_i(\bfn)
\Bigr\},
\label{N=2 lattice action}
\end{align}
where 
\begin{align}
 \Phi_1(\bfn) &= \frac{i}{2}\left(U_{14}(\bfn)-U_{41}(\bfn)+U_{23}(\bfn)-U_{32}(\bfn)\right), \nn \\
 \Phi_2(\bfn) &= \frac{i}{2}\left(U_{24}(\bfn)-U_{42}(\bfn)+U_{31}(\bfn)-U_{13}(\bfn)\right), \\
 \Phi_3(\bfn) &= \frac{i}{2}\left(U_{34}(\bfn)-U_{43}(\bfn)+U_{12}(\bfn)-U_{21}(\bfn)\right), \nn 
\end{align} 
with 
\begin{equation}
U_{\mu\nu}(\bfn) \equiv U_{\mu}(\bfn)U_\nu(\bfn+\bfe_\mu) 
U_\mu^\dagger(\bfn+\bfe_\nu) U_\nu^\dagger(\bfn),  
\end{equation}
and 
\begin{align}
 \Psi_1 &= \cL^+_{4}\psi_{(1)}(\bfn)+\cL^+_{1}\psi_{(4)}(\bfn) 
          +\cL^+_{3}\psi_{(2)}(\bfn)+\cL^+_{2}\psi_{(3)}(\bfn), \nn \\
 \Psi_2 &= \cL^+_{4}\psi_{(2)}(\bfn)+\cL^+_{2}\psi_{(4)}(\bfn) 
          +\cL^+_{1}\psi_{(3)}(\bfn)+\cL^+_{3}\psi_{(1)}(\bfn), \\
 \Psi_3 &= \cL^+_{4}\psi_{(3)}(\bfn)+\cL^+_{3}\psi_{(4)}(\bfn) 
          +\cL^+_{2}\psi_{(1)}(\bfn)+\cL^+_{1}\psi_{(2)}(\bfn), \nn 
\end{align}
with 
\begin{equation}
 \cL^+_{\nu}\psi_{(\mu)}(\bfn)\equiv \psi_{(\mu)}(\bfn)
  +U_\nu(\bfn) \psi_{(\mu)}(\bfn+\bfe_\nu) U_\nu^\dagger(\bfn).
\end{equation}
The BRST transformation is given by 
\begin{alignat}{2}
 &Q U_\mu(\bfn) = i\psi_{(\mu)}(\bfn) U_\mu(\bfn), &\quad  
 &Q \psi_{(\mu)}(\bfn) = \phi(\bfn) 
 - U_\mu(\bfn) \phi(\bfn+\bfe_\mu) U_\mu^\dagger(\bfn) 
  +i \psi_{(\mu)}(\bfn)\psi_{(\mu)}(\bfn) \nn \\
\label{N=2 lattice BRST}
 &Q\bphi(\bfn) = \eta(\bfn), & &Q\eta(\bfn) = i[\phi(\bfn),\bphi(\bfn)], \\
 &Q\vchi(\bfn) = \vH(\bfn), & &Q\vH(\bfn) = i[\phi(\bfn),\vchi(\bfn)], 
\qquad Q\phi(\bfn) =0. \nn 
\end{alignat}

Again, the obtained lattice action (\ref{N=2 lattice action}) is 
almost that of Sugino's formulation for four-dimensional $\cN=2$ 
supersymmetric Yang-Mills theory given in \cite{Sugino:2003yb}, 
and the only difference is the existence of 
the last terms of (\ref{N=2 lattice action}). 
Thus, we conclude that, as in the case of two-dimensional
$\cN=(2,2)$ supersymmetric gauge theory, 
Sugino's lattice formulation of 
four-dimensional $\cN=2$ supersymmetric Yang-Mills theory can also 
be derived from the dimensionally reduced matrix model 
by using the orbifolding prescription together with a proper sequence of
extension and truncation of the degrees of freedom. 

%We conclude this section by making a comment on the self-dual condition 
%(\ref{self-dual again}) imposed on $\chi_{(\mu\nu)}$ and $H_{(\mu\nu)}$.
%Although (\ref{self-dual again}) is 
% 

\section{Conclusion}

In this paper, we have shown that Catterall's 
lattice formulations can be understood in terms of the
orbifolding procedure. We have explicitly demonstrated this 
by a derivation of Catterall's model based on a
complexified matrix model as a mother theory. 
The symmetry of the mother theory is enhanced by this
complexification, 
and Catterall's model possesses in fact two independent BRST
symmetries. 
We have also commented on the relationship between Catterall's model 
and a variant of Sugino's lattice formulation of 
two-dimensional $\cN=(2,2)$
supersymmetric gauge theory as derived in \cite{Takimi:2007nn}. 
We have shown that we can restrict the degrees of freedom of Catterall's
model so that a linear combination of the two BRST charges, 
$\alpha Q_+ + \beta Q_-$ $(\beta\ne\pm\alpha)$, is preserved. 
The restricted theory does not depend on the values of $\alpha$ and
$\beta$ after trivial redefinitions. 
We have also applied the procedure developed in section 2 to
topologically twisted 
four-dimensional $\cN=2$ supersymmetric Yang-Mills theory. 
The lattice theory obtained in that manner is related to
Sugino's formulation \cite{Sugino:2003yb} up to the same
kind of terms that were found in the two-dimensional case.

\vspace{0.5cm}
\noindent
{\sc Acknowledgement:}~ 
We thank S.~Catterall and T.~Takimi
for useful discussions.
S.M. also acknowledges support from 
JSPS Postdoctoral Fellowship for Research Abroad. 

\bibliographystyle{JHEP}
\bibliography{refs}

\end{document}